\documentclass[letterpaper]{article}
\usepackage[preprint]{aaai2027}
\usepackage[hyphens]{url}
\usepackage{graphicx}
\usepackage{natbib}
\usepackage{booktabs}
\usepackage{array}
\usepackage{multirow}
\usepackage{amsmath}
\usepackage{amssymb}
\usepackage[table]{xcolor}
\usepackage{placeins}
\definecolor{raspblue}{RGB}{36,104,162}
\definecolor{raspgreen}{RGB}{22,133,106}
\definecolor{rasporange}{RGB}{190,105,25}
\definecolor{softgreen}{RGB}{232,246,240}
\definecolor{softblue}{RGB}{235,242,249}
\newcommand{\method}{RASP-QAOA}
\newcommand{\best}[1]{\textcolor{raspgreen}{\textbf{#1}}}
\newcommand{\second}[1]{\textcolor{rasporange}{\underline{#1}}}
\newcommand{\parten}{\ensuremath{\mathrm{PAR}_{10}}}
\begin{document}

\title{\method{}: Resource-Aware Per-Instance Selection for Exact QAOA Simulation}
\author{Chih-Chung Hsu}
\affiliations{
    Institute of Smart Industry and Green Energy,\\
    National Yang Ming Chiao Tung University, Hsinchu, Taiwan\\
    chihchung@nycu.edu.tw\\[3pt]
    {\small Code and technical supplement:
     \url{https://github.com/jesse1029/rasp-qaoa}}
}
\maketitle

\begin{abstract}
Exact QAOA simulation spans several computational representations whose useful
regions differ sharply across graph structure, circuit depth, precision, and
available memory.  Choosing only a backend name hides these differences: an
executable choice also fixes the representation, adapter, precision mode, and
memory policy.  We introduce \method{}, a per-instance selector over ten such
actions.  It first removes actions that cannot implement the requested QAOA
semantics or execution requirements, then orders the remaining actions using
instance features; actions outside learned support are handled by analytical
work estimates.  On a content-disjoint 60-request H200 evaluation,
\method{} succeeds on all \textbf{31 requests for which at least one admissible
action completes and validates}.  Within this set it reaches \textbf{27/31}
top-1 and \textbf{31/31} top-2 selection, with 1.051 geometric-mean regret.
Its failure-penalized \parten{} score is \textbf{0.0396} times that of
development-selected CUAOA (95\% interval: 0.0085--0.1644).  A separate
30-request crossover shows that graph structure changes 16 decisions and
improves the paired penalized score, while a depth-1 stump matches gradient
boosting.  The evidence supports resource-aware representation selection at
$n\leq35$, $p\leq5$, with gains driven by representation features rather than
classifier complexity.
\end{abstract}

\section{Introduction}
No single classical representation is uniformly suitable for exact QAOA
simulation.  A full-state GPU implementation is often the fastest choice while
its $2^n$ state and auxiliary arrays fit in memory.  Tensor contraction avoids
materializing the state but can become expensive as contraction width grows.
At shallow depth, a local evaluator can restrict the computation to causal
neighborhoods.  For constraint-preserving mixers, a fixed-weight representation
reduces the basis from $2^n$ states to $\binom{n}{k}$.  These alternatives do
not merely provide different implementations of the same runtime curve; they
expose different feasible regions.

This heterogeneity matters because classical simulation is central to QAOA
development.  It supports mixer design, parameter transfer, implementation
checks, and graph-family screening before quantum-device execution
~\cite{farhi2014qaoa,hadfield2019alternating,blekos2024review}.  As experiments
grow, the simulator must be chosen repeatedly for requests that vary in graph,
depth, objective, precision, and resource budget.  A fixed backend leaves useful
representations unused, whereas a runtime predictor can rank an action that
does not implement the requested semantics or cannot execute on the active
stack.

We formulate this decision at the level at which it is actually executed.  An
\emph{action} is a representation--adapter--configuration key, not a backend
name.  Two CUDA-Q actions with different precision modes, for example, have
different numerical guarantees and memory footprints.  \method{} first forms
the subset of actions compatible with the request.  It then orders
development-supported actions from instance features and uses
representation-aware work estimates when compatible actions fall outside that
support.  The two stages answer different questions: which representations may
compete, and which one should be tried first.

On a fresh 60-request H200 test, at least one admissible action completes and
validates for 31 requests; \method{} covers all 31, compared with 19 for
development-selected CUAOA and 23 for static priority.  The remaining set
contains 19 requests without a compatible portfolio action and 10 without an
observed validated completion.  The learned path supplies 19 successes and the
analytical path 12.  A separate 30-request crossover isolates overlapping
learned actions: structural features change 16 decisions and raise top-1 from
12/30 to 25/30, while a one-level rule matches gradient boosting.

Our contributions are:
\begin{itemize}
  \item a per-instance formulation in which the decision object is a complete
  executable QAOA action and compatibility is determined before performance
  ranking;
  \item \method{}, which combines request-aware compatibility, feature-based
  ordering over development-supported actions, and analytical ordering for
  compatible representations outside that support; and
  \item a component-resolved evaluation over a ten-action H200 portfolio,
  including a fresh 60-request test, a controlled structural crossover, timing
  sensitivity, and separate hardware calibration.
\end{itemize}

\section{Related Work}
\paragraph{QAOA simulation.}
QAOA simulators trade representation generality for circuit structure.
Full-state GPU engines exploit repeated layers and diagonal costs
~\cite{lykov2023qokit,stein2024cuaoa}; QTensor uses contraction
~\cite{lykov2021qtensor}; and general frameworks provide broader execution
paths~\cite{nvidia2026cudaq,qiskit2026aer,golden2023juliqaoa}.  Multi-GPU,
compressed-state, and batched systems address scale
~\cite{bayraktar2023cuquantum,zhang2025bmqsim,jiang2025bqsim}, whereas route
selectors predict cost from circuit and hardware information
~\cite{liaqat2025datacenters,bertomeu2025maestro}.  Concurrent work combines
backend selection, adaptive complex64/128 precision, memory-triggered CPU
fallback, and an adapter layer~\cite{kumaresan2026gpu}.  Our ranked object
instead fixes QAOA semantics, precision, deployment, memory, representation,
and configuration in one executable action key.

\paragraph{Per-instance algorithm selection.}
Algorithm selection maps instance information to complementary procedures
~\cite{rice1976algorithm,kotthoff2014survey,kerschke2019automated}.  Learned
portfolios couple solver selection with predicted performance
~\cite{xu2008satzilla,xu2010hydra,lindauer2015autofolio}; related work treats
generalization, censored feedback, and parallel schedules
~\cite{balcan2021generalization,tornede2022online,liu2019parallel}.  In
planning, structured representations support online selection
~\cite{ma2020online}, while interpretable features can match complex models
~\cite{ferber2022explainable}.  We order by instance only after QAOA-specific
checks define the executable actions.

\paragraph{Structure-aware quantum optimization.}
Quantum methods exploit structure at several levels: constraint-preserving
mixers restrict evolution to feasible subspaces
~\cite{hadfield2019alternating}, learned QAOA optimizers transfer information
between graph instances~\cite{khairy2020learning}, and training-free
architecture search ranks circuits with inexpensive proxies
~\cite{he2024trainingfree}.  RelOpt imposes cost-aware formal constraints on
program optimization~\cite{zhao2026relopt}.  \method{} uses graph and request
structure to select a simulator action rather than optimize parameters or
architecture.

\section{Problem Formulation}
\subsection{Requests, actions, and success}
A request is
$x=(G,H_C,p,\mathcal M,\epsilon,B,\mathcal H)$, where $G$ and $H_C$ specify
the problem, $p$ is QAOA depth, $\mathcal M$ contains mixer and constraint
semantics, $\epsilon$ is the numerical tolerance, $B$ is the request memory
budget, and $\mathcal H$ describes the active hardware and software.  An action
$a=(i,\kappa)$ combines a representation--adapter identity $i$ with a
configuration $\kappa$ that fixes precision, memory policy, and execution
flags.

Let $A_a(x)$ indicate that action $a$ satisfies the request's semantic,
precision, deployment, and memory requirements.  The candidate set is
\[
  \mathcal C(x)=\{a\in\mathcal A:A_a(x)=1\}.
\]
After execution, $R_a(x)$ records whether the action returns within the cap and
$V_a(x)$ whether the returned value passes numerical and semantic validation.
We call $Y_a(x)=A_a(x)\wedge R_a(x)\wedge V_a(x)$ a strict completion.  This
separation prevents a diagnostically useful but incompatible return from
entering policy coverage or the timing oracle.

For requests with at least one strict completion, the oracle action is
\[
  a^*(x)=\arg\min_{a:Y_a(x)=1} t_a(x),
\]
where $t_a(x)$ is the canonical action time.  The selector observes
$A_a(x)$ and request features, but not $R_a(x)$, $V_a(x)$, or $t_a(x)$.
Actions within 5\% of the fastest strict-completion time form the first
timing-equivalent group.  Top-1 means that the selected action belongs to this
group; top-2 uses the union of the first two groups.

\subsection{Why the action key matters}
The backend name is too coarse for this decision.  In the fresh portfolio,
\texttt{cudaq\_observe} denotes a cusvsim fp32 action evaluated under the
relaxed $10^{-3}$ mode.  A separate fp64 calibration action has a different
memory envelope and numerical condition and is not a fresh policy candidate.
Likewise, a fixed-weight implementation is defined jointly by its subspace
representation and a preserving mixer.  Changing any of these choices changes
both compatibility and cost.  Table~\ref{tab:portfolio} lists the complete
fresh action keys.

\begin{table*}[t]
\centering
\scriptsize
\setlength{\tabcolsep}{3.1pt}
\caption{Executable portfolio used in the fresh H200 evaluation.  Each label
fixes a representation, adapter, precision mode, and memory policy; the
complete key separates configurations with different numerical or operational
behavior.}
\label{tab:portfolio}
\resizebox{\textwidth}{!}{%
\begin{tabular}{llllll}
\toprule
Label & Representation/adapter & Fixed configuration & Label &
Representation/adapter & Fixed configuration\\
\midrule
\texttt{full\_hbm} & internal full state & complex128; precomputed cost vector &
\texttt{cuaoa\_gpu} & CUAOA full-state adapter & complex128; 36 B/amplitude\\
\texttt{full\_recompute} & internal full state & complex128; recomputed cost vector &
\texttt{qokit\_gpu} & QOKit nbcuda full state & c128 state; fp32 diagonal\\
\texttt{local\_lightcone} & causal-neighborhood evaluator & complex128; supported
local terms &
\texttt{cudaq\_observe} & CUDA-Q cusvsim full state & fp32; relaxed $10^{-3}$\\
\texttt{fixed\_weight\_subspace} & weight-$k$ subspace & c128; preserving
mixer &
\texttt{qiskit\_aer\_gpu} & Aer GPU statevector & complex128; 24 B/amplitude\\
\texttt{fixed\_weight\_full\_embedded} & full-basis embedding & c128;
preserving mixer &
\texttt{qtensor\_gpu} & QTensor contraction & unweighted MaxCut; relaxed
$10^{-3}$\\
\bottomrule
\end{tabular}}
\end{table*}

\section{\method{}}
\subsection{Compatible representation set}
\method{} first evaluates compatibility using only pre-execution information.
Circuit checks match the objective, mixer, and constraint semantics.  Precision
checks apply to the configured action rather than the backend family.
Deployment checks require the specified package and device path.  Resource
checks compare a calibrated peak-memory estimate with
$\min\{B,0.9H_{\mathrm{VRAM}}\}$.  An fp64 complex state alone requires
$16\cdot2^n$ bytes, but resident cost arrays and workspaces depend on the
adapter.  On the measured H200 stack, the full-state estimates use a 768 MiB
process base plus action-specific bytes per amplitude; QTensor uses a
calibrated contraction boundary.  The exact coefficients and invocation
settings are reported in the supplement.

\begin{figure*}[t]
\centering
\includegraphics[width=.99\textwidth]{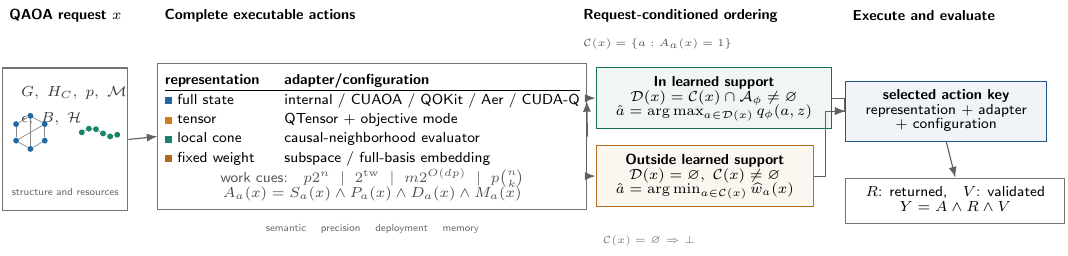}
\caption{\textbf{Decision structure of \method{}.}  The request is checked
against the semantic, precision, deployment, and memory conditions of every
complete action key.  Surviving actions form $\mathcal C(x)$.  Actions in the
fixed development support $\mathcal A_\phi$ are feature-ranked; when none
survive, representation-specific work estimates order $\mathcal C(x)$,
recovering 12/31 oracle-solvable fresh requests.  Execution is evaluated
separately by $Y=A\wedge R\wedge V$.}
\label{fig:overview}
\end{figure*}

The resulting set $\mathcal C(x)$ is representation dependent.  Dense,
moderate-size requests commonly retain full-state actions.  Sparse shallow
requests can admit local evaluation.  A preserving mixer can open a
fixed-weight route, whereas the same graph with a standard $X$ mixer cannot
use that action.  Figure~\ref{fig:overview} summarizes this separation between
compatibility and ordering.

\subsection{Feature-based and analytical ordering}
The request descriptor $z(x)\in\mathbb R^{30}$ contains graph size and density,
depth, degree statistics, fixed-weight and full-state sizes, causal-cone and
contraction proxies, objective and mixer indicators, requested precision,
VRAM, and hardware family.  Before evaluation, the learned support
$\mathcal A_\phi$ is fixed to action labels observed on the development split.
For $a\in\mathcal C(x)\cap\mathcal A_\phi$, an interchangeable ranker provides
$q_\phi(a,z(x))$.  If this intersection is nonempty, the highest-scoring action
is selected.

Compatible actions outside $\mathcal A_\phi$ are not discarded.  When
$\mathcal C(x)\neq\varnothing$ but
$\mathcal C(x)\cap\mathcal A_\phi=\varnothing$, \method{} orders the set with
representation work proxies: $p\binom{n}{k}$ for fixed-weight simulation,
$m2^{\min(24,2+dp)}$ for local evaluation,
$\tfrac14p2^{\min(n,28)}$ for tensor contraction, and full-state terms that
scale with $p2^n$ and the resident or recomputed cost representation.  These
estimates provide an ordering, not a guarantee of completion.

Writing $\mathcal D(x)=\mathcal C(x)\cap\mathcal A_\phi$, the complete policy is
\[
\hat a(x)=
\begin{cases}
\arg\max_{a\in\mathcal D(x)}q_\phi(a,z(x)),
&\mathcal D(x)\neq\varnothing,\\[2pt]
\arg\min_{a\in\mathcal C(x)}\widehat w_a(x),
&\substack{\mathcal D(x)=\varnothing\\\mathcal C(x)\neq\varnothing},\\[2pt]
\bot,&\mathcal C(x)=\varnothing,
\end{cases}
\]
where $\widehat w_a$ denotes the representation work proxy.  Learned scores
and analytical estimates are never compared on a shared scale: the
development-supported branch has priority, and the analytical branch is used
only when that support is absent from the candidate set.

All 40 development labels for the fresh evaluation are CUAOA, so
$\mathcal A_\phi=\{\texttt{cuaoa\_gpu}\}$ and the broad test does not contain a
multi-class learned boundary.  We therefore report the learned and analytical
paths separately and use a dedicated crossover to test whether structural
features improve ordering when learned-supported actions overlap.  Gradient
boosting is the default implementation; a structural stump is included to
test whether the result depends on the classifier family.

\section{Experimental Setup}
\paragraph{Workloads and execution.}
The fresh test contains 60 requests from 11 graph families, with
$n=19$--$35$, $p=1$--$5$, and sparse-local, dense-full, and
fixed-cardinality regimes.  Request identifiers, generator seeds, and circuit
content are disjoint from the 40-request development split, the earlier
benchmark, and adapter calibration.  Measurements use four NVIDIA H200 GPUs
with 143,771 MiB HBM, driver 580.95.05, CUDA 13.0, and Linux 5.15.0; every
action--request pair remains on one GPU.

Completed pairs use five independent processes.  At deployment reuse horizon
$h=8$, canonical time is the median completed
$T_{\mathrm{batch}}/8+T_{\mathrm{import}}/8$; import and batch components are
stored separately, and per-call preparation remains in $T_{\mathrm{batch}}$.
The action cap is 300 s and the worker limit 330 s.  The final data contain 366
action rows, 1,134 process invocations, and 3,402 timing measurements.

\paragraph{Validation and baselines.}
The strict evaluation uses the standard $X$ mixer for MaxCut and requires
$|v-v_{\mathrm{ref}}|\leq10^{-7}$.  A finite complex128 return selected by
fixed priority provides a provisional reference; cross-action agreement and
nine independent exact checks then validate the 31 requests entering the
oracle, with maximum error $2.84\times10^{-14}$.  Baselines are
development-selected CUAOA, static applicability priority, a compact rule
selector, the analytical ordering alone, a random-forest runtime predictor,
and supervised selector variants using the same compatibility and fallback
mechanisms.

Coverage is measured over requests with at least one observed strict
completion.  Conditional regret is
$r(x)=(t_{\hat a(x)}+t_{\mathrm{plan}})/t_{a^*(x)}$.  A failed selection receives
$10\tau$ in \parten{}, where $\tau=300$ s.  We report geometric means of
per-request \parten{} ratios; these are failure-penalized policy scores, not
kernel speedups.  Bootstrap intervals use the request as the independent unit.

\section{Results}
\subsection{Per-instance selection}
Table~\ref{tab:main_results} gives the primary H200 comparison.  \method{}
completes all 31 requests with an observed strict-completion action, reaches
27/31 top-1 and 31/31 top-2, and has 1.051 geometric-mean regret.  The median
regret is 1.0006 and the 90th percentile is 1.079.  All four top-1 misses are
successful selections in the second timing-equivalent group.

\begin{table*}[t]
\centering
\small
\setlength{\tabcolsep}{5.0pt}
\caption{Fresh H200 comparison on the 31 requests with at least one observed
strict completion.  The \parten{} column is normalized to
development-selected CUAOA; lower is better.}
\label{tab:main_results}
\begin{tabular}{lrrrrr}
\toprule
Policy & Coverage & Top-1 & Top-2 & Normalized \parten{} & Median planning\\
\midrule
Development-selected CUAOA & 19/31 & 19/31 & 19/31 & 1.000 & 7.2 $\mu$s\\
Static applicability & 23/31 & 23/31 & 23/31 & 0.308 & 10.4 $\mu$s\\
Rule selector & \best{31/31} & 27/31 & \best{31/31} &
\second{0.0408} & 15.5 $\mu$s\\
Analytical ordering & \best{31/31} & 20/31 & 29/31 & 0.0668 & 32.0 $\mu$s\\
RF runtime predictor & 28/31 & 27/31 & 28/31 & 0.145 & 37.8 ms\\
No resource conditions & 29/31 & 26/31 & -- & -- & --\\
\rowcolor{softgreen}
\textbf{\method{}} & \best{31/31} & \best{27/31} & \best{31/31} &
\best{0.0396} & 22.5 $\mu$s\\
\bottomrule
\end{tabular}
\end{table*}

\begin{figure*}[t]
\centering
\includegraphics[width=\textwidth]{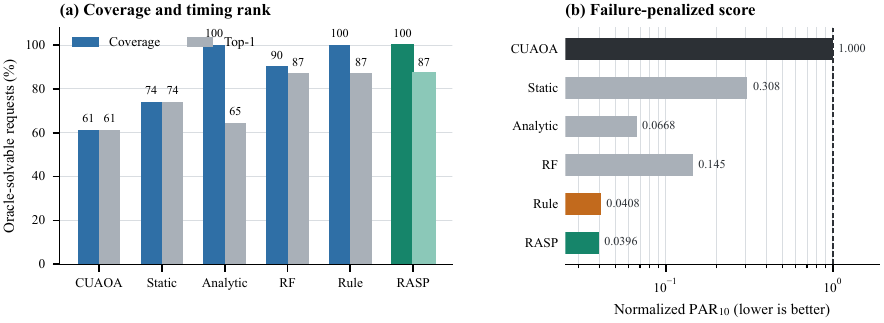}
\caption{Policy comparison on the 31 oracle-solvable fresh requests.
\textbf{(a)} Compatibility plus per-instance ordering raises coverage from
19/31 for development-selected CUAOA to 31/31, while retaining 27/31 top-1.
\textbf{(b)} Normalized \parten{} includes failure penalties and is not a
wall-clock kernel speedup.}
\label{fig:policy_summary}
\end{figure*}

Relative to development-selected CUAOA, \method{} adds 12 covered requests and
eight top-1 selections.  The paired improvements are 12--0 for coverage
($p=.00049$) and 8--0 for top-1 ($p=.0078$).  Its geometric-mean per-request
\parten{} ratio is 0.0396 (95\% bootstrap interval: 0.0085--0.1644), or a
25.26-fold lower failure-penalized score.  Against static priority the ratio is
0.1285 (0.0342--0.4253), with 8--0 coverage and 5--1 top-1 wins.

The rule selector matches \method{} in coverage and top-$k$, and its paired
penalized-score interval against \method{} contains one.  The random-forest
predictor reaches the same top-1 count but misses three solvable requests.
Analytical ordering covers all 31 yet places 11 selections outside the first
timing group.  Together, these comparisons attribute the observed gains to
compatibility-aware portfolio construction and per-instance ordering.

\subsection{Where coverage comes from}
\begin{figure*}[t]
\centering
\includegraphics[width=.69\textwidth]{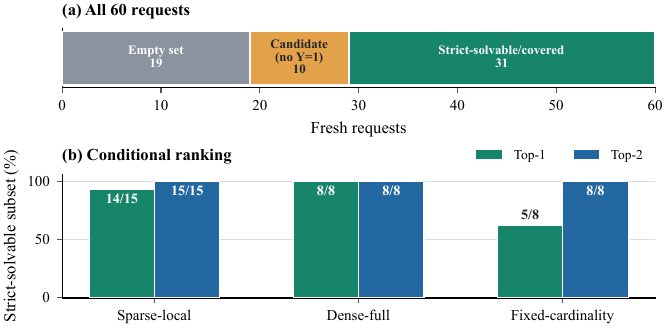}
\caption{\textbf{Portfolio reach and conditional ranking.}
\textbf{(a)} Among all 60 fresh requests, 19 have no compatible action, 10 have
one or more candidates but no strict completion, and 31 are oracle-solvable;
\method{} covers all 31.  \textbf{(b)} On this 31-request set, top-2 covers
every request in all three regimes, while top-1 is 14/15, 8/8, and 5/8.}
\label{fig:regime_summary}
\end{figure*}

Figure~\ref{fig:regime_summary} keeps the all-request context alongside the
conditional selection result.  The portfolio contains an observed strict
completion for 31 of 60 requests, and \method{} covers every one.  Nineteen
requests admit no action in the evaluated portfolio.  Ten admit at least one
action, but none completes and validates within the execution budget.  These
groups indicate missing portfolio coverage and execution limits, respectively;
neither is counted as a policy success.

The fastest-action composition explains why fixed CUAOA covers only part of
the test.  CUAOA is the sole fastest action in 19 cases.  Fixed-weight actions
appear in eight fastest sets and internal full-state actions in four; four sets
contain timing-equivalent ties.  \method{} completes every case in each group.
Three top-1 misses lie near the fixed-weight implementation crossover.  The
fourth selects a local evaluator for a sparse $n=19,p=1$ request and has the
largest regret, 3.313.

\subsection{Component attribution and structural crossover}
The development-supported path accounts for 19 strict successes.  Removing
the analytical path loses the other 12, while removing resource conditions
causes two additional coverage failures (Table~\ref{tab:main_results}).
Analytical ordering alone attains full coverage but only 20/31 top-1.  The
compatibility mechanism is therefore responsible for safe portfolio
construction, and the ordering determines timing quality within that set.

The separate structural crossover fixes $n=20,p=2$, varies ten graph families,
and gives every request at least two candidate actions.  It uses 30 development
and 30 content-disjoint test requests, with five independent processes for
each of three actions.  Full structural features complete 30/30 requests with
25/30 top-1 and a 1.029 geometric-mean \parten{} ratio.  Using only $n,p$
completes 27/30 with 12/30 top-1 and a 6.802 ratio.  Full features change 16/30
decisions and yield a paired ratio of 0.151 (95\% interval: 0.052--0.375),
with 16 wins, 14 ties, and no losses.  A depth-1 structural stump exactly
matches the gradient-boosting decisions.  Instance structure matters in this
crossover region; classifier complexity does not.

\begin{figure*}[t]
\centering
\includegraphics[width=.79\textwidth]{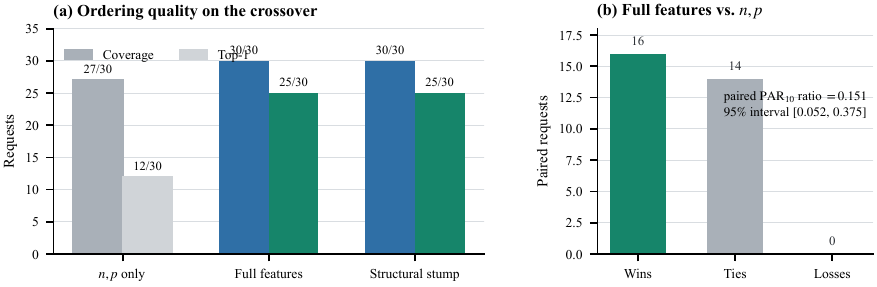}
\caption{\textbf{Structure-aware ordering on the 30-request crossover.}
\textbf{(a)} Full features raise coverage from 27/30 to 30/30 and top-1 from
12/30 to 25/30; a structural stump matches gradient boosting.
\textbf{(b)} Full features yield 16 paired wins, 14 ties, and no losses against
$n,p$-only ordering, with a 0.151 paired \parten{} ratio.}
\label{fig:crossover}
\end{figure*}

\subsection{Representation boundaries}
All four fresh top-1 misses are successful selections in the second
timing-equivalent group (Table~\ref{tab:boundary_sensitivity}a).  Three occur in the
fixed-cardinality regime, where the subspace and full-basis implementations
cross over at small $n$ and different $k$.  Their regrets are between 1.079 and
1.138.  The remaining miss is a sparse, shallow request for which the local
evaluator is valid but loses to the full-state action after end-to-end costs
are included.  Its regret of 3.313 is the largest on the fresh test.

This pattern separates ranking errors from representation failures.  The
selected action remains compatible and validated in every case, and the second
timing group contains the selection in all four.  Better modeling near the
fixed-weight crossover can improve top-1 without changing coverage, whereas
the 12 requests recovered by the analytical path require a different
representation from development-selected CUAOA.  Within-action resampling
also leaves the top-1 and top-2 counts fixed, indicating that these four labels
are stable under the observed process-to-process timing variation.

\subsection{Numerical and timing sensitivity}
Adapter checks establish the action-specific numerical scope.  CUAOA and QOKit
pass all 39 strict calibration processes; CUDA-Q fp64 passes 6/6, while its
fresh fp32 action is restricted to the relaxed mode.  QTensor is restricted to
unweighted MaxCut and excluded from the strict oracle.  Qiskit Aer passes six
of nine calibration processes; its rejection is tied to the complete request
and memory condition rather than a blanket $(n,p)$ rule.  Re-evaluating
$\epsilon\in\{10^{-7},10^{-6},10^{-5},10^{-3}\}$ leaves the headline counts
unchanged at 31 covered, 27 top-1, and 31 top-2.

Fifteen action--request rows had a single timeout that could alter the oracle
set.  Two additional independent H200 processes for each row produce
30 further timeouts, leaving the 31/19/10 partition unchanged.  Using the 680
independent completion timings, 20,000 within-action resamples give 95\% count
intervals of 27--27 for top-1 and 31--31 for top-2.

The ordering is also stable across finite deployment horizons
$h\in\{1,2,4,8,16,32\}$: coverage remains 31/31 and top-1 remains 27/31.
Cold-start timing gives 28/31 top-1; pure steady-state timing changes eight
oracle winners and reduces top-1 to 19/31.  Runtime comparisons are therefore
reported for the explicit eight-use deployment setting rather than presented
as hardware-independent rankings.

\begin{table*}[t]
\centering
\footnotesize
\renewcommand{\arraystretch}{1.08}
\caption{\textbf{Boundary, timing, and numerical robustness.}  (a) All four
top-1 misses remain successful and fall in the second 5\% timing-equivalent
group.  (b) Across reuse horizons $h=1$--$32$, coverage remains 31/31 and
top-1 remains 27/31; \parten{} is normalized to development-selected CUAOA.
(c--d) Outcome and adapter checks use the same $A/R/V/Y$ definitions as the
timing oracle.}
\label{tab:boundary_sensitivity}
\begin{minipage}[t]{.47\textwidth}
\centering
\textbf{(a) Successful selections outside the first timing group}\\[2pt]
\setlength{\tabcolsep}{2.8pt}
\begin{tabular}{lrrr}
\toprule
Request & Selected (s) & Fastest (s) & Regret\\
\midrule
$n19,k10,p5$ & 12.623 & 11.567 & 1.091\\
$n19,k5,p3$ & 0.859 & 0.796 & 1.079\\
$n23,k6,p3$ & 10.028 & 8.812 & 1.138\\
Sparse $n19,p1$ & 0.646 & 0.195 & 3.313\\
\bottomrule
\end{tabular}
\end{minipage}\hfill
\begin{minipage}[t]{.50\textwidth}
\centering
\textbf{(b) Import-amortization sensitivity}\\[2pt]
\setlength{\tabcolsep}{3.0pt}
\begin{tabular}{lrrrr}
\toprule
Timing basis & Coverage & Top-1 & Regret & \parten{}\\
\midrule
Cold start & 31/31 & 28/31 & \best{1.010} & 0.0681\\
$h=1$ & 31/31 & \best{27/31} & 1.020 & 0.0618\\
\rowcolor{softblue}$h=8$ & 31/31 & \best{27/31} & 1.051 & 0.0396\\
$h=32$ & 31/31 & \best{27/31} & 1.073 & 0.0328\\
Steady state & 31/31 & 19/31 & 1.183 & \best{0.0260}\\
\bottomrule
\end{tabular}
\end{minipage}
\par\vspace{4pt}
\begin{minipage}[t]{.42\textwidth}
\centering
\textbf{(c) Canonical action-row outcomes}\\[2pt]
\setlength{\tabcolsep}{4.2pt}
\begin{tabular}{clrr}
\toprule
Symbol & Meaning & True & Fraction\\
\midrule
$A$ & request-compatible & 148 & 40.4\%\\
$R$ & returned within cap & 192 & 52.5\%\\
$V$ & numerically validated & 124 & 33.9\%\\
\rowcolor{softgreen}$Y$ & $A\wedge R\wedge V$ & \best{107} & 29.2\%\\
\bottomrule
\end{tabular}
\end{minipage}\hfill
\begin{minipage}[t]{.55\textwidth}
\centering
\textbf{(d) H200 adapter-configuration checks}\\[2pt]
\setlength{\tabcolsep}{3.2pt}
\begin{tabular}{lrrl}
\toprule
Action & Strict & Relaxed & Evaluated mode\\
\midrule
CUAOA & \best{33/33} & 33/33 & complex128\\
QOKit & \best{6/6} & 6/6 & c128 state\\
Qiskit Aer & 6/9 & 6/9 & complex128\\
CUDA-Q fp32 & 0/3 & \best{3/3} & relaxed only\\
CUDA-Q fp64 & \best{6/6} & 6/6 & calibration only\\
QTensor & excluded & \best{3/3} & unweighted MaxCut\\
\bottomrule
\end{tabular}
\end{minipage}
\end{table*}

\subsection{Outcome separation and secondary evidence}
Compatibility, process return, and numerical validation fail for different
reasons.  Across the 366 canonical action rows, 148 actions are compatible
with their request, 192 return within the cap, 124 return a validated value,
and 107 satisfy all three conditions.  Seventeen
rows return validated values despite being incompatible with the policy
request; five compatible returns fail validation; and 36 compatible actions do
not return.  Conflating these outcomes would either enlarge the oracle with
actions the selector was not allowed to choose or hide execution failures.

The fresh H200 evaluation provides the confirmatory ten-action portfolio result,
whereas the crossover isolates structure-aware ordering.  The
earlier 120-request set is reported only as secondary evidence because 29 of
its identifiers were used during adapter calibration; under the common timing
basis it gives 69/69 coverage, 67/69 top-1, 69/69 top-2, and 1.007
geometric-mean regret.  The hardware calibration compares two internal actions,
so its evidence is limited to the stated two-action comparison.

\subsection{Hardware calibration}
\begin{figure*}[t]
\centering
\includegraphics[width=.80\textwidth]{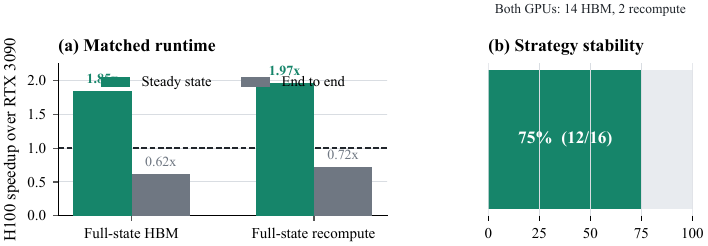}
\caption{Matched H100/RTX~3090 measurements for two internal full-state
actions.  Runtime components differ enough to require target-stack
recalibration; the preferred action agrees in 12 of 16 cases.}
\label{fig:hardware}
\end{figure*}

The two internal full-state actions expose why timing must remain tied to a
deployment setting.  On the 16 matched cases, H100 steady-state execution is
1.83--1.97 times faster than RTX~3090, whereas end-to-end ratios are
0.62--0.72 because import and setup dominate these short runs
(Figure~\ref{fig:hardware}).  The preferred action nevertheless agrees on
12/16 cases.  Representation identity transfers more often than the raw timing
score, but neither is assumed invariant across stacks.

\section{Discussion and Conclusion}
The experiments separate representation coverage from ranking quality.
Compatibility and the analytical path expand coverage from 19 to 31 requests,
while structural features improve ordering in the controlled region where
multiple actions overlap.  The broad fresh test evaluates portfolio reach,
whereas the crossover isolates timing decisions among overlapping actions.
Analytical ordering alone reaches all 31 requests but places only 20 selections
in the first timing group, compared with 27 for \method{}.  Portfolio breadth
supplies coverage, and request features refine ordering within the candidate
set.  Neither result depends on gradient boosting as a specific classifier.

The full 60-request scope also identifies concrete opportunities for a wider
portfolio.  Sparse and dense requests account for all 19 cases with no
compatible action and call for a new validated representation or broader
semantic support.  Seven of the ten candidate-without-completion cases are
fixed-cardinality requests; these cases instead expose implementation or
resource limits.  The two groups separate portfolio gaps from ordering errors.

\paragraph{Extending the portfolio.}
A new representation can be introduced without pretending that its runtime is
already learned.  Its action key specifies the adapter and configuration; its
semantic, precision, deployment, and memory conditions determine whether it
can enter $\mathcal C(x)$; and a representation work estimate places it on the
analytical path.  Only after development observations assign labels to that
action does it enter $\mathcal A_\phi$.  This staged inclusion is useful for
QAOA tooling, where a specialized simulator may have a clear execution domain
before sufficient cross-family timing data exist.  Conversely, changing the
precision or memory policy creates a new key rather than silently changing the
meaning of an existing training label.

\paragraph{Interpreting the performance numbers.}
The 0.0396 ratio summarizes failure-penalized outcomes.  Successful selections
are described separately by 1.051 geometric-mean regret and median planning
time of 22.5 $\mu$s.  Reporting both quantities distinguishes the ability to
choose a completing representation from the finer problem of matching the
fastest compatible action.  The horizon analysis further shows that coverage
can remain fixed while timing ranks shift with the inclusion of import and
setup costs.  Reuse horizons from one to 32 calls retain 31/31 coverage and
27/31 top-1, whereas steady-state timing changes eight oracle winners and
reduces top-1 to 19/31.  The ranking is tied to the declared deployment
horizon.

\paragraph{What is learned.}
The fresh development split contains one fastest supported label and tests how
a supported full-state route coexists with previously unseen compatible
representations.  The crossover supplies the complementary multi-action
comparison.  Once three actions overlap,
structure-aware features alter more than half of the test decisions and
improve both coverage and timing rank.  Because the stump and gradient booster
agree, this evidence belongs to the representation features and decision
decomposition.  Graph structure separates cases with the same $n$ and $p$ that
favor different full-state or local representations.  The ranker remains an
interchangeable component and can be replaced without changing compatibility
or analytical ordering.

\paragraph{Failure-aware scope.}
Oracle-solvable means that at least one evaluated action satisfies the fixed
request conditions and is observed to complete and validate within the cap.
The other 29 requests remain outside this observed evaluation set; a larger memory
budget, longer execution cap, or newly validated representation can move them
into it.  Keeping all 60
requests in Figure~\ref{fig:regime_summary} makes that dependence visible, while
conditioning regret and top-$k$ on the 31 requests with an observed
alternative keeps policy errors separate from portfolio gaps.  The repeated
timeout study leaves this oracle set unchanged under the reported budget.
Within it, \method{} has no coverage miss; its four top-1 misses are validated
actions in the second timing-equivalent group.

The present conclusions cover exact expectation-value simulation at
$n\leq35$, $p\leq5$ under the reported software and hardware configurations.
Broader objectives, mixers, approximate simulation methods, and target
hardware require their own action definitions, validation conditions, and
timing calibration.

\method{} turns heterogeneous exact QAOA simulation into a resource-aware
per-instance selection problem.  On the fresh H200 test it covers all 31
requests with an observed admissible strict completion, reaches 27/31 top-1
and 31/31 top-2, and obtains a 0.0396 \parten{} ratio to
development-selected CUAOA.  The structural crossover shows that request
features improve ordering when representations overlap.  The contribution is
representation-aware selection under explicit QAOA and execution requirements,
with separate compatibility, learned-support, and analytical-ordering stages.

\clearpage
\bibliography{references}

@misc{farhi2014qaoa,
  title = {A Quantum Approximate Optimization Algorithm},
  author = {Farhi, Edward and Goldstone, Jeffrey and Gutmann, Sam},
  year = {2014},
  eprint = {1411.4028},
  archivePrefix = {arXiv},
  primaryClass = {quant-ph},
  doi = {10.48550/arXiv.1411.4028}
}

@article{blekos2024review,
  title = {A Review on Quantum Approximate Optimization Algorithm and Its Variants},
  author = {Blekos, Kostas and Brand, Dean and Ceschini, Andrea and Chou, Chiao-Hui and Li, Rui-Hao and Pandya, Komal and Summer, Alessandro},
  journal = {Physics Reports},
  volume = {1068},
  pages = {1--66},
  year = {2024},
  doi = {10.1016/j.physrep.2024.03.002}
}

@article{hadfield2019alternating,
  title = {From the Quantum Approximate Optimization Algorithm to a Quantum Alternating Operator Ansatz},
  author = {Hadfield, Stuart and Wang, Zhihui and O'Gorman, Bryan and Rieffel, Eleanor G. and Venturelli, Davide and Biswas, Rupak},
  journal = {Algorithms},
  volume = {12},
  number = {2},
  pages = {34},
  year = {2019},
  doi = {10.3390/a12020034}
}

@incollection{rice1976algorithm,
  title = {The Algorithm Selection Problem},
  author = {Rice, John R.},
  booktitle = {Advances in Computers},
  volume = {15},
  pages = {65--118},
  publisher = {Academic Press},
  year = {1976},
  doi = {10.1016/S0065-2458(08)60520-3}
}

@article{kotthoff2014survey,
  title = {Algorithm Selection for Combinatorial Search Problems: A Survey},
  author = {Kotthoff, Lars},
  journal = {AI Magazine},
  volume = {35},
  number = {3},
  pages = {48--60},
  year = {2014},
  url = {https://ojs.aaai.org/aimagazine/index.php/aimagazine/article/view/2460}
}

@article{kerschke2019automated,
  title = {Automated Algorithm Selection: Survey and Perspectives},
  author = {Kerschke, Pascal and Hoos, Holger H. and Neumann, Frank and Trautmann, Heike},
  journal = {Evolutionary Computation},
  volume = {27},
  number = {1},
  pages = {3--45},
  year = {2019},
  doi = {10.1162/evco_a_00242}
}

@article{lindauer2015autofolio,
  title = {{AutoFolio}: An Automatically Configured Algorithm Selector},
  author = {Lindauer, Marius and Hoos, Holger H. and Hutter, Frank and Schaub, Torsten},
  journal = {Journal of Artificial Intelligence Research},
  volume = {53},
  pages = {745--778},
  year = {2015},
  doi = {10.1613/jair.4726}
}

@inproceedings{lykov2023qokit,
  title = {Fast Simulation of High-Depth {QAOA} Circuits},
  author = {Lykov, Danylo and Shaydulin, Ruslan and Sun, Yue and Alexeev, Yuri and Pistoia, Marco},
  booktitle = {Proceedings of the SC '23 Workshops of The International Conference on High Performance Computing, Network, Storage, and Analysis},
  pages = {1443--1451},
  publisher = {ACM},
  year = {2023},
  doi = {10.1145/3624062.3624216}
}

@inproceedings{stein2024cuaoa,
  title = {{CUAOA}: A Novel {CUDA}-Accelerated Simulation Framework for the {QAOA}},
  author = {Stein, Jonas and Blenninger, Jonas and Bucher, David and Eder, Peter J. and {\c C}etiner, Elif and Zorn, Maximilian and Linnhoff-Popien, Claudia},
  booktitle = {2024 IEEE International Conference on Quantum Computing and Engineering (QCE)},
  pages = {280--285},
  publisher = {IEEE},
  year = {2024},
  doi = {10.1109/QCE60285.2024.10292}
}

@inproceedings{lykov2021qtensor,
  title = {Performance Evaluation and Acceleration of the {QTensor} Quantum Circuit Simulator on {GPUs}},
  author = {Lykov, Danylo and Chen, Angela and Chen, Huaxuan and Keipert, Kristopher and Zhang, Zheng and Gibbs, Tom and Alexeev, Yuri},
  booktitle = {2021 IEEE/ACM Second International Workshop on Quantum Computing Software (QCS)},
  pages = {27--34},
  publisher = {IEEE},
  year = {2021},
  doi = {10.1109/QCS54837.2021.00007}
}

@inproceedings{golden2023juliqaoa,
  title = {{JuliQAOA}: Fast, Flexible {QAOA} Simulation},
  author = {Golden, John and Baertschi, Andreas and O'Malley, Dan and Pelofske, Elijah and Eidenbenz, Stephan},
  booktitle = {Proceedings of the SC '23 Workshops of The International Conference on High Performance Computing, Network, Storage, and Analysis},
  pages = {1454--1459},
  publisher = {ACM},
  year = {2023},
  doi = {10.1145/3624062.3624220}
}

@misc{nvidia2026cudaq,
  title = {{CUDA-Q} State-Vector Simulator Backends},
  author = {{NVIDIA}},
  year = {2026},
  howpublished = {\url{https://nvidia.github.io/cuda-quantum/latest/using/backends/sims/svsims.html}},
  note = {Benchmark package version 0.12.0, commit 6adf92bcda4df7465e4fe82f1c8f782ae69d8bd2; accessed 2026-07-04}
}

@misc{qiskit2026aer,
  title = {{Qiskit Aer}: Running with Threadpool and {GPU}},
  author = {{Qiskit Aer Contributors}},
  year = {2026},
  howpublished = {\url{https://qiskit.github.io/qiskit-aer/howtos/running_gpu.html}},
  note = {Benchmark package versions Qiskit 1.4.5 and Qiskit Aer 0.15.1; accessed 2026-07-04}
}

@article{xu2008satzilla,
  title = {{SATzilla}: Portfolio-Based Algorithm Selection for {SAT}},
  author = {Xu, Lin and Hutter, Frank and Hoos, Holger H. and Leyton-Brown, Kevin},
  journal = {Journal of Artificial Intelligence Research},
  volume = {32},
  pages = {565--606},
  year = {2008},
  doi = {10.1613/jair.2490}
}

@article{xu2010hydra,
  title = {{Hydra}: Automatically Configuring Algorithms for Portfolio-Based Selection},
  author = {Xu, Lin and Hoos, Holger H. and Leyton-Brown, Kevin},
  journal = {Proceedings of the AAAI Conference on Artificial Intelligence},
  volume = {24},
  number = {1},
  pages = {210--216},
  year = {2010},
  doi = {10.1609/aaai.v24i1.7565}
}

@article{liu2019parallel,
  title = {Automatic Construction of Parallel Portfolios via Explicit Instance Grouping},
  author = {Liu, Shengcai and Tang, Ke and Yao, Xin},
  journal = {Proceedings of the AAAI Conference on Artificial Intelligence},
  volume = {33},
  number = {01},
  pages = {1560--1567},
  year = {2019},
  doi = {10.1609/aaai.v33i01.33011560}
}

@article{ma2020online,
  title = {Online Planner Selection with Graph Neural Networks and Adaptive Scheduling},
  author = {Ma, Tengfei and Ferber, Patrick and Huo, Siyu and Chen, Jie and Katz, Michael},
  journal = {Proceedings of the AAAI Conference on Artificial Intelligence},
  volume = {34},
  number = {04},
  pages = {5077--5084},
  year = {2020},
  doi = {10.1609/aaai.v34i04.5949}
}

@article{balcan2021generalization,
  title = {Generalization in Portfolio-Based Algorithm Selection},
  author = {Balcan, Maria-Florina and Sandholm, Tuomas and Vitercik, Ellen},
  journal = {Proceedings of the AAAI Conference on Artificial Intelligence},
  volume = {35},
  number = {14},
  pages = {12225--12232},
  year = {2021},
  doi = {10.1609/aaai.v35i14.17451}
}

@article{tornede2022online,
  title = {Machine Learning for Online Algorithm Selection under Censored Feedback},
  author = {Tornede, Alexander and Bengs, Viktor and H{\"u}llermeier, Eyke},
  journal = {Proceedings of the AAAI Conference on Artificial Intelligence},
  volume = {36},
  number = {9},
  pages = {10370--10380},
  year = {2022},
  doi = {10.1609/aaai.v36i9.21279}
}

@article{ferber2022explainable,
  title = {Explainable Planner Selection for Classical Planning},
  author = {Ferber, Patrick and Seipp, Jendrik},
  journal = {Proceedings of the AAAI Conference on Artificial Intelligence},
  volume = {36},
  number = {9},
  pages = {9741--9749},
  year = {2022},
  doi = {10.1609/aaai.v36i9.21209}
}

@article{khairy2020learning,
  title = {Learning to Optimize Variational Quantum Circuits to Solve Combinatorial Problems},
  author = {Khairy, Sami and Shaydulin, Ruslan and Cincio, Lukasz and Alexeev, Yuri and Balaprakash, Prasanna},
  journal = {Proceedings of the AAAI Conference on Artificial Intelligence},
  volume = {34},
  number = {03},
  pages = {2367--2375},
  year = {2020},
  doi = {10.1609/aaai.v34i03.5616}
}

@article{he2024trainingfree,
  title = {Training-Free Quantum Architecture Search},
  author = {He, Zhimin and Deng, Maijie and Zheng, Shenggen and Li, Lvzhou and Situ, Haozhen},
  journal = {Proceedings of the AAAI Conference on Artificial Intelligence},
  volume = {38},
  number = {11},
  pages = {12430--12438},
  year = {2024},
  doi = {10.1609/aaai.v38i11.29135}
}

@article{zhao2026relopt,
  title = {Relational Verification for Cost-Aware Quantum Program Optimization},
  author = {Zhao, Ziming and Li, Tingting and Li, Zhaoxuan and Yin, Jianwei},
  journal = {Proceedings of the AAAI Conference on Artificial Intelligence},
  volume = {40},
  number = {17},
  pages = {14414--14422},
  year = {2026},
  doi = {10.1609/aaai.v40i17.38457}
}

@inproceedings{bayraktar2023cuquantum,
  title = {{cuQuantum SDK}: A High-Performance Library for Accelerating Quantum Science},
  author = {Bayraktar, Harun and Charara, Ali and Clark, David and Cohen, Saul and Costa, Timothy and Fang, Yao-Lung L. and Gao, Yang and Guan, Jack and Gunnels, John and Haidar, Azzam and Hehn, Andreas and Hohnerbach, Markus and Jones, Matthew and Lubowe, Tom and Lyakh, Dmitry and Morino, Shinya and Springer, Paul and Stanwyck, Sam and Terentyev, Igor and Varadhan, Satya and Wong, Jonathan and Yamaguchi, Takuma},
  booktitle = {2023 IEEE International Conference on Quantum Computing and Engineering (QCE)},
  pages = {1050--1061},
  publisher = {IEEE},
  year = {2023},
  doi = {10.1109/QCE57702.2023.00119}
}

@inproceedings{zhang2025bmqsim,
  title = {{BMQSim}: Overcoming Memory Constraints in Quantum Circuit Simulation with a High-Fidelity Compression Framework},
  author = {Zhang, Boyuan and Fang, Bo and Ye, Fanjiang and Guo, Luanzheng and Song, Fengguang and Tallent, Nathan and Tao, Dingwen},
  booktitle = {Proceedings of the 39th ACM International Conference on Supercomputing},
  pages = {689--704},
  publisher = {ACM},
  year = {2025},
  doi = {10.1145/3721145.3725747}
}

@inproceedings{jiang2025bqsim,
  title = {{BQSim}: {GPU}-Accelerated Batch Quantum Circuit Simulation Using Decision Diagram},
  author = {Jiang, Shui and Chung, Yi-Hua and Chang, Chih-Chun and Ho, Tsung-Yi and Huang, Tsung-Wei},
  booktitle = {Proceedings of the 30th ACM International Conference on Architectural Support for Programming Languages and Operating Systems, Volume 2},
  pages = {79--94},
  publisher = {ACM},
  year = {2025},
  doi = {10.1145/3676641.3715984}
}

@misc{liaqat2025datacenters,
  title = {{QAOA} in Quantum Datacenters: Parallelization, Simulation, and Orchestration},
  author = {Liaqat, Amana and Darwish, Ahmed and Roman, Adrian and DiAdamo, Stephen},
  year = {2025},
  eprint = {2503.06233},
  archivePrefix = {arXiv},
  primaryClass = {quant-ph},
  doi = {10.48550/arXiv.2503.06233}
}

@misc{bertomeu2025maestro,
  title = {Maestro: Intelligent Execution for Quantum Circuit Simulation},
  author = {Bertomeu, Oriol and Ghayas, Hamzah and Roman, Adrian and DiAdamo, Stephen},
  year = {2025},
  eprint = {2512.04216},
  archivePrefix = {arXiv},
  primaryClass = {quant-ph},
  doi = {10.48550/arXiv.2512.04216}
}

@misc{kumaresan2026gpu,
  title = {{GPU}-Accelerated Quantum Simulation: Empirical Backend Selection, Gate Fusion, and Adaptive Precision},
  author = {Kumaresan, Poornima and Muruganantham, Pavithra and Rajendran, Lakshmi and Sivasubramani, Santhosh},
  year = {2026},
  eprint = {2604.03816},
  archivePrefix = {arXiv},
  primaryClass = {quant-ph},
  doi = {10.48550/arXiv.2604.03816}
}

\end{document}